\documentclass[10pt]{article}

\usepackage{ulem}
\usepackage{epsfig}

\usepackage{amsmath}
\usepackage{amssymb}

\usepackage{graphicx}

\usepackage{cite}

\usepackage{color}

\usepackage{lscape}

\usepackage[labelfont=bf,labelsep=period]{caption}

\makeatletter
\renewcommand{\@biblabel}[1]{\quad#1.}
\makeatother

\date{}

\begin{document}

\begin{flushleft}
{\Large
\textbf{Mechanisms explaining transitions between tonic and phasic
firing in neuronal populations as predicted by a low dimensional
firing rate model}}
\\
Anca R. Radulescu$^{1,2,\ast}$

\bf{1} Department of Psychology, University of Colorado,
Boulder, CO
\\
\bf{2} Department of Biomedical Engineering, Stony Brook University School of Medicine, Stony
Brook, NY
\\
$\ast$ E-mail: radulesc@colorado.edu
\end{flushleft}

\section*{Abstract}

\noindent
Several firing patterns experimentally observed in neural populations have been successfully correlated to animal behavior. Population bursting, hereby regarded as a period of high firing rate followed by a period of quiescence, is typically observed in groups of neurons during behavior. Biophysical membrane-potential models of single cell bursting involve at least three equations. Extending such models to study the collective behavior of neural populations involves thousands of equations and can be very expensive computationally. For this reason, low dimensional population models that capture biophysical aspects of networks are needed.\\

\noindent The present paper uses a firing-rate model to study mechanisms that trigger and stop transitions between tonic and phasic population firing. These mechanisms are captured through a two-dimensional system, which can potentially be extended to include interactions between different areas of the nervous system with a small number of equations. The typical behavior of midbrain dopaminergic neurons in the rodent is  used as an example to illustrate and interpret our results.\\

\noindent The model presented here can be used as a building block to study interactions between networks of neurons. This theoretical approach may help contextualize and understand the factors involved in regulating burst firing in populations and how it may modulate distinct aspects of behavior.

\vspace{5mm}

\noindent {\bf Keywords:} midbrain, dopamine, burst firing, dynamics
regulation, bifurcations.

\clearpage
\section*{Introduction}
\label{sec:introduction}

Different populations of cells in the nervous system of many organisms display sudden, organized, and collective changes in spiking activity. Such changes in population firing involve possibly many thousands of cells. A population burst occurs when the population firing rate suddenly increases and then goes back to the basal rate. Population bursts are produced during normal behavior, but also in pathological situations~\cite{zhang2002relations}.  Population bursts are displayed in a variety of central regions of the nervous system in vertebrates (e.g., midbrain, thalamus, subiculum, hippocampus, olfactory bulb, and spinal cord) and invertebrates (ventral cord and antennal lobe in insects, stomagogastric ganglia in lobster and other crustaceans). In addition, population bursts are believed to underlie different aspects of normal and pathological function~\cite{friedman2008vta} in the nervous system. For instance, periodic bursting in the respiratory groups of the mammalian brainstem occurs at fixed phase lags~\cite{mccrimmon1997modulation, bailey2001effect}. These oscillations in population firing are also present in networks of motor neurons that control locomotion and other rhythmic activities~\cite{sqalli1993oscillatory, tierney1992physiological}. Oscillations in population activity are also important in sensory processing. For
instance, olfactory projection neurons in the antennal lobe of many insects such as moths~\cite{christensen1987male}, flies~\cite{wilson2004transformation}, locust~\cite{bazhenov2005fast}, and honeybees~\cite{krofczik2008rapid} display short-lasting responses to short-lasting olfactory stimuli. The different populations involved in these olfactory responses also display oscillatory firing for long-lasting stimuli~\cite{heinbockel1998pheromone,joerges1997representations}. Population bursts are also believed to contribute to processes related to learning and memory. For instance, pyramidal cell bursts in the hippocampus are believed to underlie the initial representation and further transference of memory traces from short term to long term storage~\cite{buzsaki1989two, wilson1994reactivation}.

There has been a considerable search for methods to appropriately study population activity, especially in neurocomputation studies related to perceptual decision making~\cite{medvedev2001synchronization,wong2006recurrent,brown2005simple,roxin2008neurobiological}, central pattern generator~\cite{buchanan1992neural,williams1992phase,varkonyi2008derivation}, synchronization~\cite{medvedev2001synchronization}. The focus of this paper is to construct a computationally efficient model to study macroscopic biophysical mechanisms underlying transitions between different kinds of population firing. The model presented here was created with the idea of studying large circuits formed by different regions of the nervous system (e.g. the hippocampus-nucleus accumbens-pallidum-VTA loop~\cite{LismanGrace2005}). One requirement for the construction of the model was that the same general formulation should be used as a template to model different populations of neurons, perhaps only differing in the choice of parameter spaces.

A variety of the currently existing network models are based on single cell activity. Some of these models include phenomenological population density formulations based on integrate and fire neurons~\cite{knight1972dynamics,knight1972relationship, knight2000approach,nykamp2000population}, poisson processes~\cite{cateau2006relation}, and generalized linear point processes~\cite{toyoizumi2009mean}. Among other limitations, these models do not include possibly important dependencies on physiologically relevant phenomena such as different sources of input with different time scales for excitation or
inhibition.  In comparison, biophysical single cell models require either two or three equations~\cite{av1993basic} (but see~\cite{izhikevich2006dynamical} for an interesting hybrid approach) and typically at least 4 parameters per ionic current. That is, biophysical single cell models are often too complex to be directly used as building blocks for a larger neural network. One problem is that the number of equations in a network model with biophysical cells is at least a two times the number of neurons, but possibly much larger. Another potential problem is that the dependence of the model dynamics on the parameters can become intractable depending on the level of heterogeneity of the cells in the model.  Examples of network
simulations based of several biophysical point neurons or complex multiple compartment neurons can be found, respectively in~\cite{rinzel1998analysis,traub1989model}, or~\cite{komendantov2002electrical}. The study of small networks of synaptically coupled cells is thus computationally expensive when the size of a network grows to a few thousand cells, even if homogeneity in the parameters is assumed.

For these reasons, we decided to construct a macroscopic model of population activity such that each of the parameters of the model represents an experimentally measurable quantity. That is, we required the model to be macroscopic but biophysical. Importantly, the model is flexible enough to potentially represent several distinct neural populations with the same general formulation. The general formulation used here can be potentially used as a building block for the study of large collections of interacting populations, thereby capturing interactions between different areas of the nervous system with a small number of equations.

\subsection*{Our approach}

We view the factors that bear upon cell populations and produce collective increases or decreases in the population firing rates as similar to those that produce spiking in a single cell. In single cells, ions that cross the membrane produce currents that change the membrane potential. Some ionic currents increase the membrane potential, whereas others contribute with a negative feedback that repolarizes the membrane. Analogously, there are factors that contribute to increase/decrease the firing rate in a population of cells. The above analogy can be observed in the firing rate model of Wilson and Cowan~\cite{wilson1972excitatory}. In addition, the phase space of the Wilson-Cowan model resembles the phase space of dynamical systems that describe the spiking activity of excitable cells~\cite{fitzhugh1961impulses, av1993basic}. In mathematical terms, such a similarity suggests topological equivalence between dynamical systems that represent single cells and populations. In view of the above remarks, we hypothesized that it should be possible to construct population models with similar trajectories and overall qualitative behaviors as single cell models, based on the premise that activity is determined by two processes: a fast one described by an amplifying variable (positive feedback), and a slower one, represented by recovery variable (negative feedback), as is the case for excitable membranes. The analog of an action potential in a population model would be a population burst. Sustained spiking in a single cell model would correspond to a sustained oscillation in firing rate in the population model. If a population displayes sustained (periodic) oscillations with a minimum rate close to the basal population rate then the oscillation can be regarded as sustained population bursting
(e.g. locomotion networks).

Our model of population activity (in an neural population which we will call $X$) is motivated by the above analogy. The parameters of the model can be directly related to macroscopic biophysical aspects of a tissue of choice. The model is capable of reproducing tonic firing, and fast, nonlinear transitions from low, to high, back to low firing that resemble the excitability (spikes) in single cells. These tonic-phasic-tonic transitions are the population bursting described above, and their periodicity can be changed by varying different parameters (see Figure \ref{fig:firingphase}). These mechanisms are captured by a two-dimensional dynamical system (i.e., two differential equations). In particular, drawing analogies from the phase plane
analysis of single cell models, our model explains the qualitative transitions in terms of what can be regarded as {\it{population excitability}}.

For a simple visual illustration: Figure \ref{fig:singlepop} shows the correspondence between the membrane potential dynamics of what would
be only one (typical) neuron in $X$ (represented by the one compartment, single cell minimal Av-Ron bursting model~\cite{av1993basic}, described in Supplementary Material, Text S1) and the dynamics of the whole population $X$. The transitions between different firing regimes in a {\it single cell} are shown in  Figure \ref{fig:singlepop}A, both as a function of time (left) and in phase space (right). The corresponding time-dependent and phase space 
representations of the {\it population} firing, as simulated by our system, are shown in \ref{fig:singlepop}B.\\

\begin{figure}[h!]
\begin{center}
FILE FOR FIGURE 1 ATTACHED
\end{center}
\caption{\small{{\bf Comparison of single cell transitions between firing and quiescence, and its analog in the population firing-rate based model \eqref{sys:dFdt}-\eqref{sys:dbdt}.} \textbf{A.} We used the single cell biophysical bursting model of AvRon et al.~\cite{av1993basic} to illustrate the time evolution of the membrane potential $V(t)$ \textbf{(left)} as well as the coupled phase-plane dynamics of the potential $V$ and the slower recovery variable $W$ \textbf{(right)}. The equations in the reference and the parameters used for the simulation are given in the first Supplementary Materials Text S1. \textbf{B.} The population model \eqref{sys:dFdt}-\eqref{sys:dbdt}  illustrates the time evolution of the firing rate $F(t)$ \textbf{(left)} as well as the trajectories in the $(b,F)$ plane \textbf{(right)}. It is notable that the single cell model employs three equations to trigger the bursts (i.e., transitions between quiescence and oscillations of the membrane potential), while our model captures the phenomenon simply as a high-low oscillation of the population firing rate, by a system of two equations. The parameters used in conjunction with our model to match the firing rate in {\bf B} with the firing rate in {\bf A} are also given in the first Supplementary Materials Text S1.}}
\label{fig:singlepop}
\end{figure}

In the rest of this article, and for the purposes of illustration, the parameters of the model are tuned to mimic the activity commonly observed in populations of midbrain dopaminergic neurons (MDNs); the values and ranges used are summarized in Table 1. The transitions between different population firing regimes will be described in terms of the population activity of MDNs.

\subsubsection*{Physiology of midbrain dopaminergic}

In general terms, {\it in vivo} rodent and primate MDNs from the ventral tegmental area (VTA, A10) and susbtantia nigra compacta (SNc, A9) display basal tonic firing rates of up to 20 Hz \textit{in vivo}~\cite{GraceBunney1983a,GraceBunney1983b,GraceBunney1983c,kiyatkin1998heterogeneity,schultz1993responses}. Subsets of
MDNs also burst at rates of up to 200 Hz in response to novel or partially unpredicted stimuli~\cite{schultz1998predictive,hyland2002firing}.  Importantly, bursting behavior in the VTA is displayed both at the single cell and population levels~\cite{GraceBunney1983c,schultz2002getting}.

Midbrain dopaminergic cells receive typical synaptic inputs from external sources, which can be net excitatory or inhibitory. Excitatory synaptic input through cholinergic and glutamatergic terminals is received from several sources including the subthalamic nucleus, the peduculo-pointine tegmentum~\cite{LismanGrace2005} and others~\cite{omelchenko2005laterodorsal,omelchenko2007glutamate,pan2005pedunculopontine}.  Inhibitory GABAergic synapses are activated by cells from within the VTA and SNc and also from basal ganglia nuclei~\cite{lodge2006laterodorsal} and other sources~\cite{floresco2003afferent}. Although for more particular distinctions and analyses it would be valuable to differentiate these as separate sources of input, in our model, for the sake of simplicity and generality, we group them together as a net extrinsic input rate.

In addition to fast excitatory and inhibitory chemical synaptic input, there is gap-junctional coupling between dopaminergic cells~\cite{GraceBunney1983c,komendantov2002electrical,vandecasteele2005electrical}. Electrical coupling is a widely observed phenomenon potentially significant in the synchronization of neuronal populations in some cases~\cite{komendantov2002electrical,perez2000gap}, and/or acting as a frequency filter in other situations~\cite{vandecasteele2005electrical}. For instance, electrical coupling may be responsible for generation and stabilization of burst firing in hippocampal networks~\cite{perez1994modulation,skinner1999bursting}. Electrical conductance between neurons in not limited to early brain development, as previously believed. Even though the high number of gap junctions in the immature brain declines rapidly during development~\cite{peinado1993extensive,rozental1998changes}, it is now known that electrical communication exists even between mature nerve cells. In the midbrain in particular, electrical connectivity decreases from 96\% to about 20\%~\cite{vandecasteele2005electrical} in a period of weeks, but the later degree of connectivity persists throughout adult life. While lacking definitive experimental evidence on the contribution of gap junctions to the regulation of burst firing in MDN, it is generally agreed that it induces similar firing between coupled cells, and thus constitutes a source of internal amplification, or recruitment within MDN populations. Other sources of internal amplification may include the inactivation of A-type channels~\cite{hoffman1997k}, NMDA receptor based excitation~\cite{zweifel2009feature} and neuromodulator input~\cite{pulver2009spike}) (by analogy with thalamocortical population bursting, synchronized mainly via inhibitory thalamic reticular nucleus input~\cite{destexhe1997synchronized}). Irrespective of the mechanism of this internal amplification, one of our model's aims is to establish whether it is a necessary factor underlying burst activity in the midbrain.

Animal studies show that firing rate of dopaminergic cells is also modulated via D2-dopaminergic receptor activation~\cite{GraceBunney1983c}~\cite{heinz2009intricacies}. This modulation results in down regulation of the firing in MDNs and can be regarded as a source of autoregulation within an MDN population. Extracellular dopamine is typically present in intervals of up to 300 milliseconds after a burst \cite{schultz2002getting, gonon1997prolonged}, which means that autoregulation of MDN firing should occur within that window of time in the absence of other influences. Studies have pointed out a strong tie between dopamine autoreceptor dynamics and expectation of reward in rodents~\cite{heinz2009intricacies}, and similarly, between midbrain autoregulatory factors and novelty-related traits in humans~\cite{zald2008midbrain}. Notably, however, midbrain (VTA and SNc) dopamine autoreceptors may be characteristic to the rodent brain, since they have not yet been found in the same regions in humans. This suggests that other dopamine mechanisms may perform in humans this down-modulatory function~\cite{meador1994differential}). 

Bursting in MDNs can be triggered through several pathways. For instance, glutamatergic and cholinergic inputs from the pedunculo-pointine tegmentum~\cite{lokwan1999stimulation} and laterodorsal tegmentum~\cite{lodge2006laterodorsal} are known to produce bursting in the VTA. In addition, glutamatergic inputs to nucleus accumbens and other striatal targets increase the tonic firing rate and burst firing in the VTA~\cite{floresco2001glutamatergic}.

These firing rate ranges and sources of synaptic input were taken into consideration to constrain our model. Analysis of the different firing regimes displayed by the model as function of parameters was then conducted. Our simulations explain mechanisms by which different transitions between tonic, bursting, sustained bursting, and high tonic firing occur in our model, in terms of quantities relevant to the MDN system.

\section*{Modeling methods}
\label{sec:methods}

\subsection*{Construction of the model}

As argued above, we consider a neuronal population $X$ whose activity as a whole can be captured by a representative firing rate $F$. The rate $F$ can be
thought of as a weighted sum of the firing rates from the cells in $X$ (see Figure \ref{fig:singlepop}), or alternatively, as the firing rate of a prototypical cell in $X$. It is assumed that $F$ changes as a function of two factors: (1) the average history of the cell's firing over a short time interval in the immediate past, and (2) the synaptic input and other modulatory influences. More specifically, as explained in the previous section, it is assumed that the input is a function of factors including intrinsic excitation (amplification), intrinsic inhibition (dampening), extrinsic excitation, and extrinsic inhibition. In the case of the MDNs in the rodent VTA, the intrinsic excitation can be thought of as resulting from a combination of gap-junctional coupling and NMDA receptor activation~\cite{GraceBunney1983c,hoffman1997k,komendantov2002electrical,zweifel2009feature,vandecasteele2005electrical}. Intrinsic dampening results from spike frequency adaptation and autoregulation by dopamine~\cite{GraceBunney1983c}. Extrinsic excitation would come from glutamatergic and cholinergic synaptic inputs. Extrinsic inhibition can result mainly from activation of GABAergic synapses from within~\cite{steffensen1998electrophysiological} and outside the VTA~\cite{floresco2003afferent}.

The inputs to $X$ are integrated by means of a response function ${\cal S}$

\begin{equation}
{\cal S}(y)=\left[ 1+ \exp \left(-k_S [y-y_S] \right) \right]^{-1},
\label{eq:responsefunction}
\end{equation}

\noindent with $k_S$ representing the gain of the response function (secs). $S$ has an increasing sigmoidal shape, with values between 0 and 1. That is, the population response saturates for large enough inputs and tends to zero as the total input rates decrease.

The population response to an incoming input $y(t)$ will trigger changes in $F$ within some delay $\tau_F$ as follows:

\begin{equation}
F(t+\tau_F)=\left[ F_{\max}- \tilde{F}(t)\right] \cdot {\cal S} [y(t)],
 \label{eq:Ftplustau}
\end{equation}

with $\tilde{F}(t) \sim F(t)$ and $F_{\max}$ representing, respectively, the average recent firing between $t$ and $t + \tau_F$ and the maximum firing rate of the cells in $X$. The first factor in Eq. \eqref{eq:Ftplustau} accounts for the history of firing, and the second for the response to new inputs. The maximal theoretical output that can be generated in the population in response to integrated input $y$ is $F=F_{\max} \cdot {\cal S}(y)$, if the population has no history of recent coordinated firing. Since delays between reception of input and response are typically very short (of the order of a few milliseconds),  $\tau_F$ can be assumed to be small. As a consequence, the discrete-time formulation in Eq. \eqref{eq:Ftplustau} can be approximated by a continuous-time equation as:

\begin{equation}
\tau_F \frac{dF}{dt}=-F(t) + [F_{\max}- F(t)] \cdot {\cal S}(y).
\label{eq:dFdt}
\end{equation}

In agreement with the analogy between cellular excitability and population excitability discussed above, we introduce a dynamic variable slower than $F$ representing negative feedback that tends to decrease the firing response to input. We achieve this by letting the intrinsic excitation and inhibition terms be, respectively, $aF$ and $b_{\max}b$, with $b \in [0,1]$ acting as a ``sliding control'' that tends to decrease the firing rate for a given input.
The dynamics of $b$ can be written as:

\begin{equation}
\tau_b \frac{db}{dt} =  b_{\infty}(F) - b   \label{eq:dbdt}
\end{equation}

with  $\tau_b$ representing the  time constant of the negative feedback, and steady state function

\begin{equation}
b_{\infty}(F) = \frac{1}{1+ \exp\left[-k_b (F-F_b) \right]}.
\label{eq:binfty}
\end{equation}

The parameter $k_b$ in Eq. \eqref{eq:binfty} represents the maximum rate of change of $b$ at steady state with respect to the firing rate $F$. The firing rate at which the steady state of $b$ is 1/2 is $F_b$. In single cell models, $b$ would be similar to the gating of potassium channels that allows repolarization of the membrane during an action potential.

The gate $b$ is not the only parameter which may change over time. However, in the present study the other parameters are kept fixed, and the dynamics of the system

\begin{eqnarray}
\tau_F \frac{dF}{dt}&=&-F-  {\cal S}(aF-b_{\max}b+P) (F-F_{\max})
\label{sys:dFdt} \\
\tau_b \frac{db}{dt}&=& b_\infty(F)-b.
\label{sys:dbdt}
\end{eqnarray}

are studied for a variety of fixed values using bifurcation analysis.

\subsection{System parameters}

\noindent Before beginning to analyze the dynamics of the system under perturbation, we note that quantitative choices for parameter values and ranges are also based on experimental evidence from MDN recordings in freely moving rodents.

\begin{table}[h] \centering
\begin{tabular}{|l|c|c|l|}
\hline
{\bf Name} & {\bf Range/Value} & {\bf Units}&{\bf Description}\\
\hline \hline
$F_{\max}$ & 400 & Hz
& Maximal firing rate of population $X$ \\
\hline $a$ & [0,1] & none & Intrinsic amplification strength\\
\hline $P$ & 0--200 & Hz & Intrinsic amplification rate\\
\hline $b_{\max}$ & 160 & Hz & Intrinsic dampening rate\\
\hline $F_b$ & 0--200 & Hz & Intrinsic dampening half activation rate \\
\hline $k_b$ & 0.025 & secs & Intrinsic dampening rate slope \\
\hline $y_S$ & 80 & Hz & Response function half maximum rate \\
\hline $k_S$ & 0.2 & secs & Response function rate slope \\
\hline $\tau_F$ & 1/400=0.0025 & sec& Population Rate Time constant\\
\hline $\tau_b$ & 1/30=0.33 & sec& Intrinsic dampening time constant\\
\hline
\end{tabular}
\caption{\small{{\bf Parameters for the model
\eqref{sys:dFdt}-\eqref{sys:dbdt}.} }} \label{table:params}
\end{table}

\noindent {\bf Integration of inputs.} In principle, we allow firing in $X$ to range between zero and $F_{\max}=400$ Hz. However, $F_{\max}$ is just a theoretical absolute maximumx, and $X$ cells spends most of their time at much lower firing rates (in fact, we show in the next section that firing is constrained to remain below an asymptotic limit of $F_{\max}/2=200$ Hz, consistent with the maximum rates found experimentally~\cite{kiyatkin1998heterogeneity}). According to known distributions of firing rates found in experimental recordings, we considered the ``critical'' firing rate (at which the cells are ``at half capacity,'' and most sensitive to input) to be $y_S=80$ Hz. The time-constant for the DA neurons, representing the entire time interval over which input can be integrated, was found to range up to $12$ msec~\cite{GraceBunney1983a}. We estimate the typical response-time of the population (time spent by $X$ on integrating incoming input before responding with an action potential) to be significantly lower, and in our analysis we considered values from $\tau_F$ from $1/800 \sec= 1.25$ msec, to $1/200 \sec = 5$ msec). 

\noindent {\bf External input.} The cumulative firing rate of external inputs to MDN can be very large (the midbrain can get net excitatory inputs of the order of 1-2kHZ). This translates post-synaptically to a much lower rate $P$, on the order of $100-200$ Hz, which is the effective rate integrated by the MDN to create an EPSP.

\noindent {\bf Intrinsic amplification.} $a \in [0,1]$ acts as a ``coupling index'', that is the fraction of the maximal electrical coupling (i.e., all cells within $X$ are mutually coupled). Dye coupling studies in midbrain regions observed that a single DA neuron may be coupled with one to five similar cells~\cite{komendantov2002electrical}. Grace and Bunney showed that 2 to 5 neurons out of the 18 SNc cells under study were electronically coupled~\cite{GraceBunney1983c}, and Vandecasteele et al. found coupling indeces from $17\%$ to $40\%$ at various development ranges~\cite{vandecasteele2005electrical}. Electrical studies found pair indeces from $20\%$ to $96\%$, and various average coupling conductances during the development~\cite{vandecasteele2005electrical}. To incorporate all theoretical possibilities, in our analysis we considered all values of $a \in [0,1]$; however, as will be seen, most plausible dynamics occur for $a \in [0.1, 0.4]$, which is consistent with the experimental findings.

\noindent {\bf Internal dampening.} Since dopamine seems to be typically present extracellularly up to 300 milliseconds after a burst~\cite{schultz2002getting, gonon1997prolonged}, we considered that the autoregulation of MDN firing by DA autoreceptor occurs within that time-frame ($\tau_b \sim 0.03 \sec$). At its maximum capacity, we considered the effect of DA receptor autoregulation to be equivalent to that of a direct inhibitory input $b_{\max}$=160 Hz. The highest receptor sensitivity $k_b$ to extracellular dopamine, corresponding to ``half-capacity,'' is obtained at the firing rate $F=F_b$. Since one of the aims of our model is to study the effects of receptor sensitivity on burst firing, we allowed $F_b$ to be a free model parameter, varying widely between $F_b=0$ and $F_b=200$ Hz.\\

Fine tuning of these parameters does not substantially affect the big picture, which remains qualitatively robust in a neighbourhood of the ranges considered. Finally, let us note that, although not all the model parameters are independent (e.g., changing $y_S$ and $P$ clearly has the same effect), we prefer to keep these quantities separate, in order to preserve their physiological meaning, and we vary only the ones that are suspected to relate to regulation of bursting activity. In the discussion, we further show that many of the dynamic properties of the model, constrained to these parameter values and ranges, agree qualitatively and quantitatively with population firing considerations drawn from electrode recordings.

\subsection*{Basic analysis and bifurcations}
\label{sec:basicanalysis}

Finding the transitions between different qualitative states of a dynamical system generally starts with plotting its nullclines (the curves in the system's phase-plane where either $\frac{dF}{dt}=0$ or $\frac{db}{dt}=0$), then calculating the equilibria $(b_*,F_*)$ of the system (i.e., the points where the two derivatives are simultaneously zero). In fact -- as the Av-Ron single compartment cell burster~\cite{av1993basic}, and the original Wilson-Cowan model~\cite{wilson1972excitatory} -- our model falls in the class of Fitzhugh-Nagumo systems, which pioneered geometrical analysis of phase-portraits in the context of neurocomputational models~\cite{fitzhugh1961impulses}. While the general properties of the FitzHugh-Nagumo system have been studied and discussed at lengh~\cite{medvedev2001synchronization}, we will hereby focus on the quantitative phase plane characteristics, and on the parameter transitions more particular to our model.

In our model, a first easy remark concerns the region of the plane which contains the relevant dynamics. Indeed, start by noticing that, for any input $y$ and for any choice of parameters, the response ${\cal S}(y) \in (0,1)$. Since $dF/dt=-F+(F_{\max}-F){\cal S}(y)$, this implies that $dF/dt<0$ for $F \geq F_{\max}$ and that $dF/dt>0$ for $F \leq 0$. Hence, if $0<F<F_{\max}$, then $-F<dF/dt<F_{\max}-2F$, so that $dF/dt<0$ when $F>F_{\max}/2$. So all trajectories end asymptotically in the open strip $0<F<F_{\max}/2=200$. In addition to this, one can also note that we always have $0<b_{\infty}(F)<1$, which in turn implies that $db/dt<0$ for $b \geq 1$ and that $db/dt>0$ for $b \leq 0$, constraining the long term dynamics to the open strip $0<b<1$. In conclusion: the open region ${\cal R}=(0,1) \times (0,200)$ traps the asymptotic phase plane dynamics of system, in the sense that all trajectories eventually end in ${\cal R}$ and that any attracting sets (hence any features relevant to the long term dynamics like equilibria and cycles) are also contained in ${\cal R}$.

The high nonlinearity of the system makes exact computation of the equilibria very difficult; however, it is not hard to show that the the monotonicity of the nullclines implies existence of one up to three intersection points (true in general for any FitzHugh-Nagumo system of ODE). The stability of these points changes with their position in the phase plane, so that different set of parameters can deliver different combinations for the stability of equilibria.

While the $db/dt=0$ nullcline has sigmoidal shape, so is globally increasing, the piecewise monotonicity of the nullcline $dF/dt=0$ depends on the choice of parameters. It is its temporary slope reversal which gives rise to the possibility of multiple equilibria, hysteresis, cycles. (Equivalent conditions on the parameters for this to occur are hard to calculate, but Wilson and Cowan, for example, compute a sufficient condition in their original paper~\cite{wilson1972excitatory}.) Obtaining any exact analytical information on the position and geometry of the cycles, when they exist, is even more intractable (as one would generally expect for this degree of nonlinearity). The fact that both nullclines are contained in the region ${\cal R}$ implies, of course, that any phase plane cycle would have to also be contained in ${\cal R}$, which is a useful initial estimate. Some simulation-based illustrations of equilibria and cycles, and their possible coexistence in the phase plane are plotted in Figure \ref{fig:firingphase}, together with representative trajectories.\\

\begin{figure}[h!]
\begin{center}
FILE FOR FIGURE 2 ATTACHED
\end{center}
\caption{\small{{\bf Different firing behaviors and phase plane configurations.} The left and right columns show, respectively, the firing rate $F$ vs. time and the $(b,F)$ phase plane. \textbf{A1-2.} The system has an attracting fixed point at $F_\infty$=33.9137 Hz,$b_\infty$=0.3425. Parameters: $F_b$=60 Hz, $a=0.1$, $P=120$ Hz, $b_{\max}$=160 Hz. \textbf{B1-2.} Sustained oscillations. Same parameters as in \textbf{A}, except $a$=0.2. \textbf{C1-2.} Low tonic firing (green) and bursting (blue) in a bistable regime. Parameters: $a$=0.5, $F_b$=28.131 Hz. Initial conditions: for the tonic firing $F_0$=40 Hz, $b_0$=0.4; for the bursting $F_0$=40 Hz, $b_0$=0.7. Stable fixed point at $F_{\infty}$=11.1343 Hz and $b_{\infty}$=0.3950. \textbf{D1-2.} Bursting (blue) and high tonic firing (green) in a bistable regime. Parameters: $a$=0.5, $F_b$=139.98734. Stable fixed point at $F_\infty \sim$ 189 and $b_\infty$=0.7737. Initial conditions: for the tonic firing $F_0$=40, $b_0$=0.4; for the bursting $F_0$=40, $b_0$=0.7.}}
\label{fig:firingphase}
\end{figure}

A bifurcation in the dynamics of a system is by definition a (parameter) state of the system where these dynamics exhibit a qualitative change. For example, for a 2-dimensional continuous time system like \eqref{sys:dFdt}-\eqref{sys:dbdt}, such a qualitative transition could be a change in the stability of its invariant sets: equilibria or cycles. A bifurcation diagram is a graph that shows these invariant sets, and possibly their type and stability, as a function of a single parameter (for co-dimension 1) or of two parameters varied simultaneously (for co-dimension 2)~\cite{guckenheimer1990nonlinear}~\cite{kuznetsov1995elements}. The stability of equilibria can be monitored computationally by following the nature (real or complex) and the sign of the local Jacobian's eigenvalues; bifurcations characterized by a local change in stability of cycles are generally harder to establish.

In our study, we use bifurcation diagrams of co-dimensions 1 and 2 to assess the qualitative differences in firing rates and transitions between such regimes -- resulting from varying the physiological parameters in the model. For illustrating and understanding these bifurcations, we use numerical computations~\footnote{The computations of the steady states and phase plane cycles as well as the bifurcation diagrams were performed in Matcont (version 2.4)~\cite{dhooge2003matcont}, a bifurcation finding software based on continuation algorithms.}; these are discussed throughout the remainder of the paper.

The stability-changing local bifurcations of codimension one that occur in our system are of three types: subcritical Hopf, limit point (or saddle node bifurcation of equilibria) and fold (saddle node bifurcation of cycles)\footnote{The terminology in the field is not always consistent between authors; throughout the paper, we remained faithful to the system of terms used by the authors of the Matcont software. For clarification, all terms used are defined in the following paragraph.}; each type of transitions corresponds to a different ``crash'', or ``degeneracy'' of the respective equilibria or cycles.

A subcritical Hopf bifurcation appears generally when an equilibrium with complex conjugate eigenvalues changes stability (as one parameter of choice is varied), so that its Jacobian's eigenvalues transition from having positive real part (equilibrium is an unstable spiral) to negative real part (stable spiral), through a pure imaginary stage. This transition also creates a repelling cycle around the stable spiral, whose radius increases locally with the distance to the bifurcation point (see Supplementary Material, S2). A limit point (or saddle node) bifurcation appears when two equilibria (one stable and one unstable) traveling in the phase-plane collide and disappear, through an intermediate stage of a common half-stable equilibrium (i.e, with one null eigenvalue of the Jacobian). Similarly, in a fold bifurcation, two nested cycles of opposite stabilities collide and disappear, through a bifurcation stage of a common half-stable cycle. For illustrations of successions of such transitions, see Figure \ref{fig3:tonicbursting}, or the Supplementary Material, Figures S2 and S3.

When varying simultaneously two parameters, we encounter two types of codimension 2 bifurcations~\cite{guckenheimer1990nonlinear}. The presence of the hysteresis in the phase-plane opens the possibility for {\it{cusp bifurcations}}; these occur where two branches of limit point bifurcation curve meet tangentially. For neighboring parameter values, the system has three equilibria which collide and disappear pairwise via limit point bifurcations. A Bogdanov-Takens codimension 2 bifurcation appears generically at the intersection of a limit-point curve, a Hopf curve and a homoclinic curve\footnote{For nearby parameter values, the system has two equilibria (exactly one of which is a saddle) which collide and disappear via the limit-point bifurcation. The nonsaddle equilibrium undergoes a subcritical Hopf bifurcation generating an unstable cycle. This cycle degenerates into an orbit homoclinic to the saddle and disappears via a saddle homoclinic bifurcation.}. Although interesting in the context of the system's dynamics, the parameter ranges seem to force the physiological dynamics to remain away from the Bogdanov-Takens bifurcations, which is why they will not be discussed here in any further detail.



\section*{Results}
\label{sec:results}

\subsection*{Tonic versus phasic firing}

The main focus of this paper is on mechanisms that trigger and stop bursting, and how these mechanisms can be explained by changes in parameters in the model. In other words, such mechanisms will be studied by characterizing the bifurcation structure of the system. The geometry of the bursts, their timing and their kinetics can be very different, depending on the parameter values. Different ways in which parameters tune different features of the bursts are discussed below.

Figure \ref{fig:firingphase} illustrates four possible firing regimes achieved by the system \eqref{sys:dFdt}-\eqref{sys:dbdt} with four different sets of parameters. The time evolution of the firing rate $F(t)$ (Figure \ref{fig:firingphase} left column), is paired in each case with the corresponding trajectory in the $(b,F)$ phase-plane (Figure \ref{fig:firingphase}, right column). Depending on the parameter values, phase-plane trajectories can
exhibit two qualitatively different types of long-term behavior, determined by the presence or absence of two types of attractors. One such attractor is a tonic firing rate that corresponds to a stable equilibrium in the model. The other attractor is an oscillation, which corresponds to periodic transitions between low and high population firing rates; this regime may be regarded as periodic bursting of the cells in $X$. Figure \ref{fig:firingphase}A shows an example in which the system has a global stable equilibrium at $F_\infty \sim$ 33.9 Hz. When initialized at any other value, the firing may undergo an upturn, but always returns to the steady tonic rate $F_\infty$. For the parameter values in Figure \ref{fig:firingphase}A, repeated bursting is not possible. In contrast, for the parameters in Figure \ref{fig:firingphase}B, any trajectory in the $(b,F)$-plane starts cycling. The firing rate  describes an oscillation between $\sim$ 200 and $\sim$ 0 Hz. These oscillations can be thought of as a bursting regime with no tonic firing. However, tonic firing and bursting are not mutually exclusive in this model. Figures \ref{fig:firingphase}C and \ref{fig:firingphase}D illustrate the coexistence of tonic firing and bursting. The right panel of Figure \ref{fig:firingphase}C shows a stable equilibrium at $(b_\infty,F_\infty)$ and a large stable cycle. The basins of attraction of these two attractors are separated by an unstable cycle (not shown). For instance, two nearby initial states could be situated, respectively, inside and outside the region enclosed by the unstable limit cycle. The trajectory of the point inside the unstable limit cycle would go toward $F_\infty \sim$ 11 Hz. The trajectory of the point outside the unstable cycle goes  toward the limit cycle. That is, the system can asymptotically stabilize to either a low tonic firing, or toward bursting. A similar situation is pictured in Figure \ref{fig:firingphase}D, except that the stable equilibrium corresponds to high tonic rate of about $F_\infty \sim$ 189 Hz and the coexisting oscillations have longer peaks; note that the bursts are longer and the
intervals between population bursts are shorter.\\

\begin{figure}[h!]
\begin{center}
FILE FOR FIGURE 3 ATTACHED
\end{center}
\caption{\small{{\bf Examples of two successions of local bifurcations.} \textbf{A.} The steady state of the system migrates in the phase plane $(b,F)$, describing a continuous curve as $F_b$ increases from 0 Hz to 200 Hz. The stability of the equilibria is color coded as follows: blue for stable spiral, red for unstable spiral. The two changes in stability are thus noticeable along the curve (red stars) and are caused by two subcritical fold-Hopf bifurcations in the dynamics (the fold bifurcations are represented by the purple vertical lines). Cycles emerge at these bifurcations. The evolution of the large stable cycles as $F_b$ increases is represented by the blue dotted curve (described by its highest and lowest $F$ values). Since the bistability windows between the fold and the Hopf bifurcations are extremely small, we inserted them as two magnified frames; in each of these frames, the fold bifurcation is visible as a purple vertical line, and tops and bottoms of the repelling cycle as $F_b$ increases are drawn as a red curve. In panel {\bf A}, we fixed $a$=0.5 and $P$=120 Hz.
\textbf{B.} The steady state curve shows two fold bifurcations (purple vertical lines), two Hopf bifurcations (red stars), and two limit point -- or saddle node -- bifurcations (light green stars). In this case $a=0.75$, $P=100$ Hz, and $F_b$ increasing from 0 Hz to 200 Hz. Additional explanations of the transitions in the dynamics and illustrations of the phase planes for these transitions are presented in Supplementary Materials Figure S3. Common fixed parameters for {\bf A} and {\bf B}: $k_b$=0.025 secs, $k_f$=0.2 secs, $y_S$=80 Hz, $b_{\max}$=160 Hz, $\tau_F$=1/400 secs, $\tau_b$=1/30 secs.}}
\label{fig3:tonicbursting}
\end{figure}

Bifurcation diagrams were constructed to investigate the transitions from tonic to oscillatory firing and back. As an example, two bifurcation diagrams for the half-maximal rate of dampening, $F_b$, are compared in Figure \ref{fig3:tonicbursting}. The two panels in Figure \ref{fig3:tonicbursting} illustrate how the equilibria of the system change as $F_b$ increases between 0 to 200 Hz. In Figure \ref{fig3:tonicbursting}A, the amplification parameter was set to $a$=0.5. The stable and unstable rates are shown, respectively, in blue and red. In this example, a limit cycle oscillation around the unstable fixed point emerges as $F_b$ increases toward 30 Hz and disappears as $F_b$ approaches $\sim$ 140 Hz. The firing rates in the oscillatory regime alternate between $\sim$ 200 Hz (top dotted lines) and $\sim$ 0 Hz (lower dotted lines). The transition from low tonic firing into the bursting regime takes place through a thin bistability window. In this window, a stable equilibrium and the stable cycle coexist, suggesting that cells in $X$ can end up in either state, based on their recent history of firing. This coexistence has been observed experimentally in midbrain electrode recordings (for example by Grace et al.~\cite{lodge2005hippocampus,lodge2007aberrant}). The minimum and maximum rates of the repelling cycle that separates the stable equilibrium and the limit cycle are shown in the bottom inset as a red curve. The bistability onsets with the formation of a cycle around the stable state (fold bifurcation), and disappears when the unstable cycle becomes infinitesimally small and renders the equilibrium unstable (subcritical Hopf bifurcation). In the same figure, as $F_b$ continues to increase, the bursting eventually transitions sharply into high tonic firing. This transition occurs at the end of another small bistability window (top inset) formed between a subcritical Hopf and a fold bifurcations (see also Supplementary Material -- Figure S2 for a more detailed illustration of the phase plane transitions). Note that, strictly speaking, Hopf points mark the entry into the purely cycling regime, with no stable steady-state. Since the bistability windows are small, and since the Hopf points are easily identifiable on the graph of the steady state, for the remainder of this paper we will consider the Hopf points to be the mark for the onset of sustained (periodic) oscillations.

In Figure \ref{fig3:tonicbursting}B, we show the analogous bifurcation diagram for a higher intrinsic amplification $a$=0.75. The transition from low-tonic firing to bursting is qualitatively the same as in Figure \ref{fig3:tonicbursting}A, i.e., through a fold-Hopf bifurcation and a subsequent bistability window. The transition from bursting to high tonic firing is, however, more mathematically complex, and involves the appearance and disappearance of two additional fixed points, through two limit-point bifurcations (see also Supplementary Materials Figure S3). Although of a slightly different nature than the one in Figure \ref{fig3:tonicbursting}A, this transition produces the same final result, as oscillations cease for the larger rates of activation for intrinsic dampening.

\subsection*{Triggering sustained firing rate oscillations (tonic-bursting regime)}

We study first how our population responds to extrinsic excitation with and without the intrinsic amplification, respectively, for $a=0$ and $a>0$ (Figure \ref{fig4:trigbursting}A-B).  Each curve corresponds to a different value of $F_b$. The steady state $F_{\infty}$ of the population firing rate increases, as the frequency of extrinsic excitatory input, $P$, increases from 0 to 200 Hz.\\

\begin{figure}[h!]
\begin{center}
FILE FOR FIGURE 4 ATTACHED
\end{center}
\caption{\small{{\bf Steady state firing rate as a function of extrinsic input $P$.} \textbf{A.} No intrinsic amplification ($a$=0), and increasing levels of dampening $F_b$. From lower to higher curve: $F_b=20$ Hz (red curve), $F_b$=50 Hz (blue curve), $F_b$=100 Hz (green curve), and $F_b$=150 Hz (orange curve). There are no Hopf points on any of these curves, hence periodic bursting can't be triggered for $a$=0. \textbf{B.} In the presence of sufficient intrinsic amplification, an extrinsic excitation appropriately balancing the dampening will trigger periodic bursting. Along these curves, Hopf points are marked with red stars, and limit points are marked with light green stars. The steady states are unstable (spirals or saddles) between the Hopf points, and stable otherwise (stability not shown). To maintain clarity of the diagram, and since their approximate placement is clear, the fold bifurcations and the corresponding cycles are not shown. Parameters: $F_b$=100 Hz, and increasing levels of intrinsic amplification. From lower to higher curve: $a=0$ (green curve, identical with the green curve in \textbf{A}), $a$=0.1 (purple curve), $a$=0.25 (cyan curve), $a=0.5$ (orange curve) and $a=0.75$ (blue curve). Fixed parameters common to {\bf A} and {\bf B}: $k_b$=0.025 secs, $\tau_b$=0.03 secs, $k_f$=0.2 secs, $y_S$=80 Hz, $P$=100 Hz, $b_{\max}$=160 Hz, $\tau_F$=0.0025 secs.}}
\label{fig4:trigbursting}
\end{figure}

As shown in Figure \ref{fig4:trigbursting}A, a unique, stable, and increasing steady state rate $F_{\infty}$ as a function of level of excitation $P$ occurs when there is no intrinsic dampening. In other words, the system has only one stable state, and no attracting cycles. That is, periodic bursting cannot be obtained if no amplification is present. The steady state rate $F_{\infty}$ increases as the dampening onset decreases (i.e., a larger $F_b$ rate places $F_{\infty}$ on a higher curve). That is, the model naturally predicts that the (asymptotic) population firing rate will be higher if dampening starts at higher rates.  The values of $P$ for which $F$ clearly departs from 0 depend on the half-activation of dampening, $F_b$. If onset of dampening occurs at lower rates (red curve in the bottom of Figure \ref{fig4:trigbursting}), the input firing rate required to increase the population firing rate is much larger than if dampening occurs at higher rates.

To trigger oscillations, it is necessary for the system to have a nonzero intrinsic amplification (Figure \ref{fig4:trigbursting}B). However, although
necessary, having $a>0$ is not sufficient to produce sustained bursting. As shown in Figure \ref{fig4:trigbursting}B, the onset of firing rate oscillations characteristic of periodic bursting also requires a large enough net balance between the extrinsic inputs and the onset of intrinsic dampening. In addition, bursting requires initial build-up to some low level of tonic firing to pass the stability threshold into the cycling regime. Such thresholds, located at the Hopf points, are shown as red stars in Figure \ref{fig4:trigbursting}B.

\subsection*{Modulating phasic firing}

In general, the duration of high firing rates of the oscillatory regime can be modulated by intrinsic excitation and dampening (Figures \ref{fig:BurstDur_a} and \ref{fig:BurstDur_Fb}), and extrinsic input (Figure \ref{fig:burstfreq}). First, consider the changes in burst duration as a function of the intrinsic amplification weight $a$.  The weight $a$ modulates the length of the bursts (Figure \ref{fig:BurstDur_a}A-B). The larger $a$,
the longer the bursts, with interburst intervals relatively constant. The transition from high/low firing rate oscillations into high tonic
firing occurs for a sufficiently large increase in $a$ (see also Figures \ref{fig:BurstDur_a}B  \ref{fig4:trigbursting} and \ref{fig:BifCod1}B).\\

\begin{figure}[h!]
\begin{center}
FILE FOR FIGURE 5 ATTACHED
\end{center}
\caption{\small{{\bf Increasing the duration of the high-firing phase of the oscillation without changing the low-firing phase.} The high firing rate portion of the oscillation (burst) can be increased by increasing the weight of the intrinsic amplification $a$. The figure shows the dependence on $a$ of: the brust intervals (red curve), the inter-burst intervals (green curve) and their sum, i.e., the period of the oscillation (blue curve). As the weight of intrinsic amplification increases, the length of the bursts is longer, but the relaxation intervals between bursts remain approximately unchanged. Fixed parameter values: $k_S$=0.2, $k_b$=0.025, $y_S$=80 Hz, $F_b$=60 Hz, $P$=120 Hz, $b_{\max}$=160 Hz, $\tau_F$=2.5 msec, $\tau_b$=33 msec.}}
\label{fig:BurstDur_a}
\end{figure}

Other effects can be obtained by decreasing the intrinsic dampening activation rate (i.e., increasing $F_b$) as sustained oscillations take place. Figure \ref{fig:BurstDur_Fb} shows how the activation of intrinsic dampening and the extrinsic input can vary so that the high firing rate portion of the cycle is longer and the quiescence intervals shorter, and without changing the firing frequency (not shown). This effect is important, as it suggests a mechanism by which populations of bursting cells can regulate their duty cycles (burst duration/cycle duration) without altering their bursting frequency.\\

\begin{figure}[h!]
\begin{center}
FILE FOR FIGURE 6 ATTACHED
\end{center}
\caption{\small{{\bf Changing the duty cycle without changing the oscillation frequency.} Increasing the intrinsic amplification rate $P$ ({\bf A}) or the half activation rate of dampening $F_b$ ({\bf B}) increases the duration of the high/state of the firing rate oscillation, without changing the oscillation frequency. The result is an increase in the duty cycle of the population. Fixed parameter values: $k_S$=0.2, $k_b$=0.025, $y_S$=80 Hz, $b_{\max}$=160 Hz, $a$=0.5, $\tau_F$=2.5 msec, $\tau_b$=33 msec. For {\bf A}, $P=100$ Hz; for {\bf B}, $F_b=60$ Hz.}}
\label{fig:BurstDur_Fb}
\end{figure}

The frequency of the sustained oscillations in the model can be also controlled by varying the time constants $\tau_F$ and $\tau_b$. These parameters can be used to regulate the length of the burst and inter-burst intervals (Figure \ref{fig:burstfreq}A), and the symmetry of the rising and falling phases (Figure \ref{fig:burstfreq}B). One way to think about the effect of increasing $\tau_F$ is that the right hand side of equation \eqref{sys:dFdt} will increase for smaller values of $\tau_F$, yielding larger changes in $F$ per unit time during an oscillation. On the other hand, increasing the time constant of the recovery variable, $\tau_b$, decreases the time-dependent change in $b$, which results in slower decrease during the oscillation in $F$. As a consequence, the frequency of oscillations decreases, without significantly altering the duty cycle (i.e. the burst / inter-burst duration ratio).\\

\begin{figure}[h!]
\begin{center}
FILE FOR FIGURE 7 ATTACHED
\end{center}
\caption{\small{{\bf Changing the oscillation frequency  without changing the duty cycle.} ({\bf A.}) When increasing the time constant $\tau_b$, both burst and inter-burst intervals lengthen (dotted graphs), although the duty cycle remains relatively stable. Under changes of the time constant $\tau_F$, the duration of and between bursts remains fairly stable (solid graphs). However, increasing $\tau_F$ affects the symmetry of the burst geometry, i.e., the duration of the rising phase and decaying phase of the firing rate ({\bf B}). {\bf B.} $\tau_F$=1.25 msec, $\tau_b$=33 msec (red curve) and $\tau_F$=5 msec, $\tau_b$=33 msec. Fixed parameter values: $k_S$=0.2, $k_b$=0.025, $y_S$=80 Hz, $F_b$=60 Hz, $P$=120 Hz, $b_{\max}$=160 Hz, $a$=0.5.}}
\label{fig:burstfreq}
\end{figure}

\subsection*{Ending sustained oscillations}

Similarly to the transition from equilibrium (tonic firing) into sustained firing rate oscillations, the converse transition (from oscillations to tonic firing) corresponds to a qualitative change in the dynamics of the system. To study these transitions, the attracting states of the system were calculated by varying only one parameter at a time. Two examples of these codimension 1 bifurcations are shown in Figure \ref{fig:BifCod1}. Panel A illustrates the bifurcation diagrams for the rate of extrinsic input $P$, for different values of the half activation of intrinsic dampening, ($F_b$ =80, 100, 120, and 150 Hz, shown, respectively, in blue, purple orange, and green). For instance, the lowest of the curves (green) in Figure \ref{fig:BifCod1}A shows a monotonic increase in the steady state firing as a function of $P$, for a relatively high onset of intrinsic dampening ($F_b=150$ Hz). The transition (Hopf) points of the diagram are shown by the red stars. For contrast, the first curve from left to right (blue) was calculated assuming an earlier onset of intrinsic dampening ($F_b=80$ Hz). Note that the curve is not monotonic, and two additional limit points (green stars) appear between the Hopf points. Recall that limit point (or saddle node) bifurcations are the states of the system where, under variation of one parameter, two equilibria collide and disappear, or, conversely, two equilibria emerge. For the blue curve in Figure \ref{fig:BifCod1}A, corresponding to $F_b$=80 Hz, a saddle and a repeller appear  for large enough $P$ at the first  limit point bifurcation (top green star), as new equilibria to a system already having an attracting equilibrium in the low tonic firing range. Under further increase of $P$, this low tonic equilibrium changes stability through a Hopf bifurcation (bottom red star), then steadily approaches the newly created saddle point, and eventually collides with it and vanishes at the second limit point bifurcation (bottom green star). The surviving (unstable) equilibrium will later undergo another Hopf bifurcation (top red star) and become attracting, causing the seizure-like, high tonic firing observable for very high values of $P$. The additional two equilibria are unstable, so they do not constitute attracting states for the firing in the
system. However, they play a very important role in the transitions necessary to lead from the low tonic to the high tonic firing. This succession of transitions is mathematically more complex than the fold-Hopf pair observed along the lower (monotonic) curves in Figure \ref{fig:BifCod1}A.\\

\begin{figure}[h]
\begin{center}
FILE FOR FIGURE 8 ATTACHED
\end{center}
\caption{\small{{\bf Bifurcation diagrams illustrating the stability of the steady-state firing rate as a function of one parameter.} \textbf{A.} The curves represent the locus of the steady-state firing rate as $P$ was varied from $P$=0 Hz to $P$=200 Hz, for fixed $a=0.5$. Each curve corresponds to a different value of $F_b$; from bottom to top: $F_b$=80 Hz (green curve), $F_b$=100 Hz (orange), $F_b$=120 Hz (purple), $F_b$=150 Hz (blue). \textbf{B.} The curves show the steady-state as $a$ is varied from $a$=0 to $a$=1, for fixed $P$=120 Hz; from bottom to top: $F_b$=40 Hz (purple), $F_b$=80 Hz (green), $F_b$=120 Hz (orange).  Hopf points are marked with red stars, and limit points are marked with light green stars. The steady states are unstable between the Hopf points, and stable otherwise (stability not shown). The fold bifurcations and the corresponding cycles are not shown. Fixed parameters common to {\bf A} and {\bf B}: $k_S$=0.2, $k_b$=0.025, $y_S$=80, $b_{\max}$=160, $\tau_F$=2.5 msec, $\tau_b$=33 msec.}}
\label{fig:BifCod1}
\end{figure}

Panel B in Figure \ref{fig:BifCod1} shows the bifurcation diagram for the amplification weight $a$, using different values of the dampening half activation ($F_b$ = 40, 80, and 120 Hz, respectively, in orange, green, and purple). Observe that the steady state firing rate $F_{\infty}$ is noticeably higher for higher values of $F_b$, and also, limit points are present for a larger range of $F_b$.

The cessation of bursting doesn't necessarily occur symmetrically with the bursting onset, and may involve a different succession of bifurcations (compared to the equilibrium-to-cycle transition).  In principle, there are multiple and qualitatively different mechanisms of transition from bursting back into tonic firing, each corresponding to different parameter regimes. As has been observed in different experimental settings, the system can transition from bursting into either low or high tonic firing.

\subsubsection*{From oscillations into high tonic firing}

Our first phase-plane and bifurcation plots, staring with Figure \ref{fig:firingphase}, suggested that cycling is possible for specific parameter interval. Figure \ref{fig3:tonicbursting} showed that there is an interval of $F_b$ for which the system exhibits cycling, when all other parameters are held fixed. Along each such curve, there are two Hopf points -- the approximate marks for the system entering and leaving the bursting mode. As noted earlier, increasing either type of excitation (i.e., increasing $P$ in Figure \ref{fig:BurstDur_a}A or increasing $a$ in Figure \ref{fig:BurstDur_a}B), or decreasing the intrinsic dampening (i.e., increasing $F_b$ in Figure \ref{fig3:tonicbursting}) leads the system down a path from low tonic firing through an oscillation window, and  into a regime of high tonic firing. However, it is not unreasonable to assume that the predicted seizure-like plateau after the end of oscillations (of $\sim 200$ Hz) would be impossible to sustain in real neural populations, eventually leading to cellular death~\cite{fergestad2008neuropathology, meller2003seizure}(see Discussion Section).

\subsubsection*{Transitions into low tonic firing}

Sustained firing rate oscillations can be efficiently stopped by shifting the onset of dampening towards lower firing rates. One way in which neuron populations may experience this transition, could be a decrease in the activation of dampening, shortly after entering the bursting mode. In MDNs, such a decrease could be triggered by a sudden increase in the local concentration of dopamine due to recruitment of cells into bursting mode. Shifting the dampening onset to lower firing rates can result into a regime where bursting is no longer possible (even though the extrinsic inputs that originally produced this bursting may still be present). The effects of this change can be visualized in Figures \ref{fig:BifCod1}A-B as ``pushing the system down'' to a lower curve (corresponding to lower $F_b$). If this drop is significant enough, the system's position on the new curve will be to the left of the Hopf point which marks the onset of bursting, in the range where the equilibrium is stable. Hence the system is forced to return to low tonic firing. It is possible that this forcing inhibitory term may be in effect for as long as necessary to prevent further bursting. For example, lowering the excitation $P$ may allow the return of $F_b$ to the pre-oscillatory range, without any need for additional protection against bursting. Conversely, raising $P$ may require even higher inhibition (lowering the curve even more), if the return of bursting is not yet desirable in the system. This suggests a negative feedback modulation through intrinsic dampening, by a continuous readjustment of $F_b$ according to the current state of the population firing (as further discussed in the last section).

Bifurcation diagrams in a two-dimensional parameter plane were used to study the global changes in the dynamics as two parameters change simultaneously and independently. In the figures in Supplementary Materials Text S4, we plotted the parameter planes $(P,a)$, $(P,F_b)$ and $(a,F_b)$. In all three plots, we notice that the Hopf curve (i.e., the curve that contains all the Hopf points for the particular ranges of parameters) delimits a closed region. This is the region where ``periodic bursting is possible'' in the system.  Recall that a Hopf point is the approximate mark for entering the cycling regime when one parameter is changed. Here, the Hopf curve is an approximate boundary for the 2-parameter region that permits periodic oscillations.

Some of these mechanisms can be more easily captured by allowing multiple system parameters to vary at the same time. Examples of the bifurcations displayed by the system as two parameters are varied simultaneously are shown in Supplementary Materials Text S4.

\section*{Discussion}

The previous paragraphs describe a model that approximates the activity of a homogeneous population of neurons. The model exhibits transitions between bursting and tonic firing that resemble the typical behavior of several populations of cells in the nervous system. We proceed with some comments on features particular to the model, which we place within the context of MDN activity. We finish the discussion with more general remarks about potential applications of the model in studying the collective behavior of different populations of cells in the nervous system.\\

In systems without electrical coupling or other intrinsic amplification, solitary bursts might appear simply as a perturbation of the initial condition due to increase in external stimulus (for example, in thalamocortical neurons, population bursting occurs typically during sleep or absence seizures and is synchronized mainly via common inhibitory inputs from the reticular nucleus). However, since MDNs in rodent are known to be partially electrically coupled (i.e., our model can be assumed to have $a>0$), the system has the potential for sustained (periodic) bursting. We believe that most of the dynamics observed in the MDN occur close to (on either side) of the fold-Hopf bifurcation, in proximity of the low bistability window, rendering singular or periodic bursts with rates up to $F=200$ Hz, which will then have to be readily suppressed when necessary (e.g., when the triggering stimulus loses its novelty content).

Some experimental studies observed typical burst rates of $20-30$ Hz~\cite{grace1984control,hyland2002firing}; in other studies, the heterogeneity of types, localization and behavior of midbrain cells transpired as wider distributions, ranging from ($20-30$ Hz) to very high (up to $150-200$ Hz)~\cite{kiyatkin1998heterogeneity}. Similarly, some spike counts quote $2-5$ spikes per burst~\cite{hyland2002firing}, with relatively small interspike intervals of $6-12$ msec~\cite{kiyatkin1998heterogeneity}; however, more comprehensive descriptions have included higher quotes, and have additionally noticed that glutamate enhances bursting, increasing spike/burst counts to $8-10$, while leaving inter-spike intervals relatively unchanged. Overall, accounts of burst duration have ranged widely from $20-120$ msec~\cite{kiyatkin1998heterogeneity,hyland2002firing}, and inter-burst intervals from $50-250$ msec~\cite{hyland2002firing}, which are generally accommodated by the range of rhythms encompassed by our model in the neighbourhood of the low bistability window.

Intrinsic amplification is necessary but not sufficient to produce periodic bursting. As for solitary bursts, the transition into oscillations has to be produced by an additional factor, such as increase in the excitatory synaptic input (e.g., the pedunculo-pontine tegmentum (PPTg) gates bursting in MDNs~\cite{pan2005pedunculopontine}), or a decrease in the dampening (e.g., a reduction in the GABAergic input to VTA neurons has been correlated to bursting activity within the VTA~\cite{steffensen1998electrophysiological}). The modeling results presented here, and interpreted in the context of MDN activity, also agree with recordings obtained from slice preparations in which current injection triggers MDN bursting only if an NMDA-receptor agonist is added to the standard saline
\cite{johnson1992burst}.

The behavior of the system \eqref{sys:dFdt}-\eqref{sys:dbdt} can be used to test the hypothesis that MDN firing states are independently regulated by two distinct afferent pathways, whose interactions control tonic and phasic activity in MDNs. On one hand, the mesolimbic dopamine system is strongly influenced by the hippocampus~\cite{LismanGrace2005, floresco2001glutamatergic}. In brief, infusions of NMDA into the ventral subiculum (vSub) increases the number of spontaneously active DA neurons (population activity), while having no effect on firing rate or average bursting activity. In
contrast, NMDA activation of the pedunculopontine tegmentum resulted in a significant increase in DA neuron burst firing without significantly affecting population activity. Simultaneous excitation of the vSub and PPTg induced significant increases in population activity and burst firing in MDNs ~\cite{floresco2003afferent}~\cite{lodge2006laterodorsal}. As already discussed, extrinsic and intrinsic excitation have different effects on population firing. In addition, a more detailed study of the influence of extrinsic inputs to populations of MDNs could be performed by extending the model to include more than one area, by coupling different populations like the one described by \eqref{sys:dFdt}-\eqref{sys:dbdt}. Such coupling could incorporate specific inhibitory/excitatory pathways that are well known in these circuits, but whose functional influence on firing in MDN are still not well understood. The specific interactions between cells participating in feedback loops from/to the midbrain are only speculative at this point. In sum, a potential general attribute of the model presented here is the ability to verify or suggest theoretical alternatives to existing hypotheses about the network dynamics in such loops.

The transition into oscillations in the system \eqref{sys:dFdt}-\eqref{sys:dbdt} cannot occur if the initial firing rate is low. As shown in Figure \ref{fig4:trigbursting}B, the firing activity has to first build up to a level of tonic firing of $2-10$Hz, for bursting to occur, which agrees well with experimental evidence~\cite{lodge2006laterodorsal}. Once the sustained oscillations are triggered in the model, the duration of the high-frequency phases and those of the intervals between them can be tuned by a variety of factors. In the oscillatory regime, high (low) firing rate episodes get shorter/longer when the system is brought close to any point of transition into low (high) tonic firing. In contrast, high (low)  firing rate episodes and get longer when the system is closer to a transition into high (low) tonic firing (eg. seizure-like plateau).

The oscillations in the system \eqref{sys:dFdt}-\eqref{sys:dbdt} can be stopped by two different mechanisms corresponding to a forced transition into either low tonic or high tonic firing. As noted earlier, the transition between low-tonic to oscillations resemble the busting experimentally observed in MDN populations. The accumulation of dopamine in the vicinity of bursting cells has been shown to affect MDN bursting \cite{GraceBunney1983c}. The model predicts that directly lowering the excitatory input or increasing inhibition may also stop oscillations. This prediction can be rephrased as a hypothesis that has been partially confirmed: decreasing the tone of PPT input, or increasing GABAergic tone from basal ganglia nuclei can downregulate bursting in MDNs \cite{pan2005pedunculopontine}.

The amplification or the dampening can thus be regarded as factors gating the bursting, regulating the numbers of tonic versus bursting MDNs.  More generally, the model predicts that a balance between excitation and inhibition is also required for extrinsic input to trigger firing rate oscillations. Some limitations and possible generalizations of the model \eqref{sys:dFdt}-\eqref{sys:dbdt} are discussed next.

The functional details and parameter ranges used to illustrate the behaviour of the model are specifically applicable to populations of MDNs. Therefore, the model \eqref{sys:dFdt}-\eqref{sys:dbdt} is applicable to at least VTA-like physiology and connectivity. However, given a region of interest in the nervous system, equations \eqref{sys:dFdt}-\eqref{sys:dbdt} could be set up to model  different families of cells by choosing different parameter sets, thus enabling the study of larger systems. As noted before, some of the observed qualitative features could be applicable to networks that govern population bursting in other brain regions (e.g., the thalamus~\cite{he2002differential}, or the subiculum~\cite{van2006different}).

The relationship between the population activity in  $X$ and its inputs was modeled by means of sigmoidal functions. This approach was chosen to capture the saturation effect that occurs in cells as their excitatory and inhibitory inputs are integrated \cite{wilsonetal2004}. The qualitative use of such functions in our context is justified by models and experiments that go back to Cowan, Boltzmann, and Hill~\cite{wilson1972excitatory,brozovic2008mechanism,marreiros2008population}. Nevertheless, it is worth noticing that the parameters $k_S$ and $k_b$ are only partially justified from a macroscopic perspective, at least for the MDN populations we discuss here.

The transitions between equilibrium, oscillatory and back have been studied from a static perspective. In other words, it is necessary to force the change
in parameters by hand to trigger these transitions in the current model. Future work will address this issue by incorporating time-dependent changes in the parameters of the system. By analogy with the single cell models by Av-Ron {\it{et al.}} \cite{av1993basic}, a particular bifurcation parameter
could be transformed into a new variable in the system that triggers the start and stop of  oscillations.

\subsection*{General considerations}
\label{subsec:genconsiderations}

The focus of this paper has been placed on studying qualitative changes in the system dynamics that result from  parameter variations. Most of the analysis
presented here involved variations of one-parameter at a time, without trying to achieve an overall picture of how these changes combine to determine the global behavior of the system. However, the more parameters are considered simultaneously, the harder it is to visually illustrate the results. Such a study would be very interesting and relevant to understanding the underlying mechanisms that produce these dynamics. One of the immediate goals of future work is to build upon our existing model of bursting in MDN as the center-piece of an entire network as proposed by the Lodge-Grace experiments
\cite{lodge2006laterodorsal}, with network feedback loops captured by  new equations. The theoretical framework used here potentially enables the possibility to produce testable hypotheses about mechanisms underlying  the normal and pathological activities of networks involving the hippocampal formation, basal ganglia and midbrain monoaminergic nuclei.

The local and even global phenomena observed in the model \eqref{sys:dFdt}-\eqref{sys:dbdt} have been pointed out in single cell models
\cite{av1991minimal, av1993basic, terada2000two, izhikevich2006dynamical}. The similarity between the dynamics of population firing described here, and the membrane potential dynamics in single cells, is more substantial than phase plane analogies for occasional parameter values and is currently under rigorous examination. One possible direction to follow in this regard could be to use such phase-portrait analysis of dynamics to understand the effects of dysfunction in the dopaminergic and other monoaminergic systems, which have long been thought of as primary factors in mental illnesses like anxiety, major depression or schizophrenia, and also, in pathologies like Parkinson's disease \cite{romo1992role}. This constitutes an incentive to try to contextualize and understand the factors involved in regulating monoaminergic modulation systems, and study the ways in which these modulatory networks affect other networks
~\cite{zweifel2009feature}. There have been some attempts to study the effects of dopaminergic modulation using the rat as an animal model,
~\cite{lodge2006laterodorsal, lodge2007aberrant}, and also in the clinical setting ~\cite{tremblay2005functional, nestler2006mesolimbic, sil2008role}. However, crucial information is yet missing for establishing a sustainable etiology of mental illnesses. A working translational model such as \eqref{sys:dFdt}-\eqref{sys:dbdt} may be the tool that would optimally combine clinical and basic research results.

\section*{Acknowledgments}
The author wishes to gratefully thank Marco Herrera for the invaluable consultations, as well as Erin McKiernan and Tom Hazy for all the useful suggestions and stimulating discussions.

\bibliography{FRbiblio}


\topmargin 0.0cm
\oddsidemargin 0.5cm
\evensidemargin 0.5cm
\textwidth 16cm
\textheight 22cm

\clearpage
\thispagestyle{empty}
\vspace{-2cm}
\subsection*{Supplementary Material -- Text S1}

\noindent To produce the {\bf membrane potential illustration of single cell bursting in Figure 1A}, we have used the basic biophysical model for bursting neurons proposed by AvRon et al.~\cite{av1993basic}, based on adaptation of the K channel conductance:

\begin{eqnarray}
&& C_m \frac{dV}{dt} = I - I_{Na} - I_{K} - I_{L}\\
&& \frac{dW}{dt} = \frac{W_{\infty}(V) - W}{\tau_{w}(V)} \nonumber\\
&& \frac{d\bar{g}_{K}}{dt} = S(V)(V-V_{rest})-d(\bar{g}_{K} - \bar{g}_{K}^{(rest)}) \nonumber
\end{eqnarray}

\noindent where

\begin{eqnarray}
&& I_{Na} = \bar{g}_{Na} m_{\infty}^{3}(V)(1-W)(V-V_{Na}) \nonumber\\
&& I_{K} = \bar{g}_{K}(W/s)^4(V-V_K) \nonumber\\
&& I_{L} = \bar{g}_{L}(V-V_L) \nonumber
\end{eqnarray}

\noindent and

\begin{eqnarray}
W_{\infty}(V) &=& \left( 1+\exp[-2a^{(w)}(V-V_{1/2}^{(w)})] \right)^{-1} \nonumber\\
m_{\infty}(V) &=& \left( 1+\exp[-2a^{(m)}(V-V_{1/2}^{(m)}] \right)^{-1} \nonumber\\
\tau_{w}(V) &=& \bar{\lambda} \left(  \exp[a^{(w)}(V-V_{1/2}^{(w)})]+\exp[-a^{(w)}(V-V_{1/2}^{(w)})] \right) \nonumber\\
S(V) &=& S \cdot \left( 1+\exp[-2a^{(s)}(V-V_{1/2}^{(s)})] \right) \nonumber
\end{eqnarray}

\noindent We used the following parameter values: $V_{Na}=55$ mV, $V_{K}=-72$ mV, $V_L=-49.3$ mV, $V_{rest}=-56$ mV, $\bar{g}_{Na}=120$ mS/cm$^2$, $\bar{g}_{L}=0.3$ mS/cm$^2$, $\bar{g}_{K}^{(rest)}=8$ mS/cm$^2$, $C_m=1$ $\mu$F/cm$^2$, $s=1$, $a^{(w)}=0.045$, $V_{1/2}^{(w)}=-55$ mV, $a^{(m)}=0.05$, $V_{1/2}^{(m)}=-33$ mV, $a^{(s)}=0.2$, $V_{1/2}^{(s)}=-20$ mV, $\lambda=0.5$, $S=0.04$, $p=0.008$.

\vspace{1cm}

\noindent To produce the {\bf mean field firing rate illustration of a bursting neural population in Figure 1B}, we used the two-dimensional model described in this paper:

\begin{eqnarray}
&& \tau_F \frac{dF}{dt} = -F+ [F_{\max}-F]{\cal S}[aF-b_{\max}b+P] \\
&& \tau_b \frac{db}{dt} = b_\infty(F)-b \nonumber
\end{eqnarray}

\noindent where

\begin{eqnarray}
&& {\cal S}(Y)=\frac{1}{1+\exp[-k_S(y-y_S)]} \nonumber\\
&& b_{\infty}(F) = \frac{1}{1+ \exp\left[-k_b (F-F_b) \right]} \nonumber
\end{eqnarray}

\noindent We used the following parameter values: $F_{\max}=400$ Hz, $a=0.685$, $P=100$, $b_{\max}=160$, $\tau_F=2.5$ msec, $\tau_b=66$ msec, $k_S=0.2$, $y_S=80$ Hz, $k_b=0.025$, $F_b=60$ Hz.

\clearpage
\thispagestyle{empty}
\oddsidemargin 0.0cm
\topmargin -.5cm
\textwidth 20cm
\begin{landscape}

\subsection*{Supplementary Material -- Figure S2 (FIGURE FILE ATTACHED)}
\begin{figure}[h!]
\begin{center}
\end{center}
\caption{{\small {\bf A succession of four qualitative transitions (bifurcations) undergone by the system as one parameter is varied.} In this case, $F_b$ was increased from $0$ to $200$ Hz, while the other system parameters were held fixed (in particular, the intrinsic amplification was fixed to $a=0.5$; compare with Figure 3A in the main text). We illustrate the main features of the phase-plane for each interval between two consecutive bifurcations. For small values of $F_b$, there is a globally attracting equilibrium, situated in the low firing range ({\bf A}). Increasing $F_b$ past the first critical value, makes the system undergo a fold bifurcation: one half-stable cycle appears around the equilibrium. Under further increases in $F_b$, this cycle separates into a large attracting cycle (blue cycle in {\bf B}), and a small repelling cycle (red cycle in {\bf B}). The unstable cycle is the boundary between the basins of attraction of the stable equilibrium and of the stable cycle. Continuing to increase $F_b$ maintains for a short while this bistability regime, with no more qualitative changes; however, quantitatively, while the large stable cycle only changes shape slightly under increments of $F_b$, the small unstable cycle gradually shrinks around the stable equilibrium. At the next critical value for $F_b$, the unstable cycle becomes so small that is swallows up the equilibrium and disappears (subcritical Hopf bifurcation), changing the stability of the equilibrium in the process. The system has now an unstable equilibrium surrounded by a large stable cycle ({\bf C}). Under further increments of $F_b$, the unstable equilibrium will travel inside the cycle from its lower left, corresponding to low firing rates, towards the upper right, corresponding to high firing rates ({\bf D}). Reaching the third critical value of $F_b$ produces another (reverse) subcritical Hopf bifurcation, in which the small repelling cycle  reappears around the equilibrium (although now in a different position), while changing again the stability of the equilibrium ({\bf E}). Increasing $F_b$ past this point allows the unstable cycle to enlarge, and eventually collide with the surrounding stable cycle, such that they both disappear -- through a fourth bifurcation (a second fold bifurcation). The only invariant feature remaining is the stable equilibrium, now situated in the high firing range ({\bf F}).}}
\end{figure}



\subsection*{Supplementary Material -- Figure S3 (FIGURE FILE ATTACHED)}
\begin{figure}[h!]
\begin{center}
\end{center}
\caption{{\small {\bf A succession of six qualitative transitions undergone by the system as one parameter is varied.} Here, $F_b$ is increased from $0$ to $200$ Hz, while the other parameters are held fixed (in particular, $a=0.75$; compare with Figure 3B in the main text). As before, we illustrate the main features of the phase-plane for each interval between two consecutive bifurcations. For small values of $F_b$, analogously with Figure 1, there is a globally attracting equilibrium, situated in the low firing range ({\bf A}). Increasing $F_b$ through the next two critical values, makes the system undergo a similar fold bifurcation ({\bf B}), then a similar subcritical Hopf bifurcation ({\bf C}) as in Figure 1B . From this point on, however, increasing $F_b$ produces a succession of bifurcations significantly different from the case of $a=0.5$. At the third critical value for $F_b$, two additional (both unstable) equilibria appear inside the large stable cycle: an unstable spiral and a saddle. This critical point is a limit-point (or, by a different terminology, a saddle node) bifurcation ({\bf D}). The fourth critical value of $F_b$ brings a second subcritical Hopf bifurcation, in which a small repelling cycle  appears around the unstable spiral, changing it into a locally stable spiral ({\bf E}) and bounding its attraction basin. Increasing $F_b$ past this point allows the unstable cycle to enlarge, and eventually collide into the surrounding stable cycle (before it could crash into the nearby saddle point), such that both cycles disappear; this constitutes the fifth bifurcation -- fold bifurcation, slightly more complex than the analogous one in Figure 1B ({\bf F}). The three equilibria (a stable spiral, a saddle and an unstable spiral) all survive this transition. Increasing $F_b$ further, however, causes the two unstable equilibria to collide and disappear (the sixth, limit point, bifurcation). Past this last critical value of $F_b$, only the high firing stable spiral remains ({\bf G}).}}
\end{figure}
\end{landscape}

\clearpage
\thispagestyle{empty}
\subsection*{Supplementary Material -- Text S4}

Examples of the bifurcations displayed by the system as two parameters are varied simultaneously are shown in Figures \ref{fig:Bif1Cod2} and \ref{fig:Bif2Cod2}. Figure \ref{fig:Bif1Cod2} illustrates the bifurcation curves of the system as the extrinsic input rate $P$ and its weight $a$ are varied. Figure \ref{fig4:trigbursting}B shows the Hopf limit point transitions that occur on each steady state curve (i.e., for each distinct value of $a$), as the parameter $P$ is increased. This means that, if $a$ is varied, the Hopf points will describe a {\it{Hopf curve}}, and the limit points will describe a {\it{limit point curve}}. In Figure \ref{fig:Bif1Cod2}A, these are shown in red (the Hopf curve) and green (the limit point curve); compare with Figure \ref{fig4:trigbursting}B, where a few of these Hopf points are plotted as red stars, and a few of the limit points are plotted as green stars. The bifurcation curves are drawn over the whole parameter domain, to provide a complete illustration of the underlying mathematics, but the portions corresponding to unbiological values of $a$ (i.e., $a \notin [0,1]$) are dotted. A clearer interpretation can be formulated if one observes the same curves plotted in the
two-dimensional $(P,a)$ parameter plane, in panel B of the same figure. The red Hopf curve encloses a compact region ${\cal K}$ in this parameter plane, which corresponds to the bursting regime. Triggering bursting corresponds to a parameter path which enters ${\cal K}$, while cessation of bursting is delivered by the parameter path leaving ${\cal K}$. Similar features are found in the $(P,F_b)$ and $(a,F_b)$ parameter planes, as shown in Figure
\ref{fig:Bif2Cod2}, A and B.\\

\begin{figure}[h!]
\begin{center}
FILE FOR FIGURE 11 ATTACHED
\end{center}
\caption{\small{\textbf{Simultaneous variation  of two parameters.} \textbf{A.} Bifurcation diagram showing the evolutions of the positions of limit points and of the Hopf points as both parameters $a$ and $P$ change. The limit curve is drown in green, and the Hopf curve is in red. The curves are dotted outside of the $[0,1]$ interval for $a$. The stars mark codimension 2 bifurcations: Bogdanov-Takens (purple star), and cusp (dark green stars). See also Figure \ref{fig4:trigbursting}B. \textbf{ B.} The same two curves are illustrated in the parameter plane $(P,a)$. Fixed parameters: $k_S$=0.2, $k_b$=0.025, $y_S$=80, $b_{\max}$=160, $\tau_F$=2.5 msec, $\tau_b$=33 msec.}}
\label{fig:Bif1Cod2}
\end{figure}

\begin{figure}[h!]
\begin{center}
FILE FOR 12 ATTACHED
\end{center}
\caption{\small{\textbf{ Bifurcation diagrams illustrating the bursting regions in two-dimensional parameter planes.} \textbf{ A.} Parameter plane $(P,F_b)$. \textbf{B.} Parameter plane $(a,F_b)$. Fixed parameters: $k_S$=0.2, $k_b$=0.025, $y_S$=80, $b_{\max}$=160, $\tau_F$=2.5 msec, $\tau_b$=33 msec.}}
\label{fig:Bif2Cod2}
\end{figure}

\end{document}